\documentclass[]{article}

\usepackage[final]{style_neurips_2019_ml4ps}  
\usepackage[utf8]{inputenc} 
\usepackage[T1]{fontenc}    
\usepackage{hyperref}       
\usepackage{url}            
\usepackage{booktabs}       
\usepackage{amsfonts}       
\usepackage{nicefrac}       
\usepackage{microtype}      
\usepackage{comment} 
\usepackage{graphicx}
\usepackage{wrapfig}
\usepackage{epstopdf}

\begin{document}

\graphicspath{ {./images/} }
\epstopdfDeclareGraphicsRule{.pdf}{png}{.png}{convert #1 \OutputFile} 
\DeclareGraphicsExtensions{.png,.pdf}

\title{Unsupervised Distribution Learning for Lunar Surface Anomaly Detection}

\author{Adam Lesnikowski\\
NVIDIA\\
2701 San Tomas Expressway\\
Santa Clara, CA 95051 \\
\texttt{alesnikowski@nvidia.com} 
\And
Valentin T. Bickel\\
ETH Zurich \& MPS Goettingen\\
Sonneggstrasse 5\\
Zurich, 8092, CH\\
\texttt{valentin.bickel@erdw.ethz.ch}
\And
Daniel Angerhausen\\ 
(a) Center for Space and Habitability\\
University of Bern\\
Gesellschaftsstrasse 6\\
3012 Bern\\
\texttt{daniel.angerhausen@csh.unibe.ch}\\
(b) Blue Marble Space Institute of Science\\
1001 4th Ave, Suite 3201 \\
Seattle, Washington 98154 
}


\maketitle
\begin{abstract}

In this work we show that modern data-driven machine learning techniques can be successfully applied on lunar surface remote sensing data to learn, in an unsupervised way, sufficiently good representations of the data distribution to enable lunar technosignature and anomaly detection.
In particular we train an unsupervised distribution learning model to find the landing module of the Apollo 15 landing site in a testing dataset, with no dataset specific model or hyperparameter tuning. 
Sufficiently good unsupervised data density estimation techniques have the potential to enable a dazzling number of useful downstream tasks, including locating lunar resources for future space flight and colonization, finding new impact craters or lunar surface reshaping, and deciding the importance of unlabeled samples to send back from power- and bandwidth-constrained missions. 
We show in this work that such unsupervised learning can be successfully done in the lunar remote sensing and space sciences contexts. 

\end{abstract}

\section{Introduction, Motivations}

The search for so called technosignatures in our Solar System and beyond has gained new attention recently [1]. Here, technosignatures are physical properties or effects that provide scientific evidence of past or present extraterrestrial technology [2]. According to NASA the search for technosignatures should be performed in parallel to the search for biosignatures in the field of astrobiology, as technosignatures could reveal the existence of intelligent life elsewhere in the universe [1]. One area in the field of technosignature research is the search for non-terrestrial artifacts in our Solar System, particularly on the surfaces of planets and moons [e.g., 1, 3, 4]. [5] suggested to start a systematic search for ground-based technosignatures on the lunar surface. Besides technosignatures, there are other objects of scientific relevance on the lunar surface, such as fresh impact craters, etc. However, traditional search methods involve inspection of remote sensing data by human operators, which is a slow, biased, and inefficient process. In addition, the physical shape, form, and material of technosignatures is completely unknown and signatures could easily be missed by operators. However, there is a wealth of remote sensing data available, meaning that the bottleneck of technosignature search is the human component.

In recent years, numerous powerful machine learning methods have been developed, such as vision methods [7] and explicit data distribution learners, including variational autoencoders (VAEs) [8, 9], flow based models [10], and auto-regressive models [11]. Our motivation is to utilize such machine learning-driven approaches to automate technosignature search and to prove the potential of explicit data distribution learners to extract scientifically relevant results from space exploration data. For technosignature search, a data and ML-driven approach would provide unique advantages: 1) large data sources enable us to scan the surfaces of entire planets and moons, 2) machine learning allows us to remove the human in the loop and to automate the scanning process, and 3) the utilization of unsupervised methods allows us to scan for anomalies without a priori knowledge about the physical state and form of the to-be-detected targets. For this proof of concept, the Moon is an ideal test bed, as there is a wealth of satellite data available and as numerous exploration missions have left non-natural relics on the surface, such as the Apollo and Luna missions, that can serve as ground truth for model validation and testing. 

\section{Methods}

\subsection{Data}

The detection of objects with unknown shapes and sizes on the surface of a planet or moon requires imagery with two main characteristics, 1) sufficient spatial resolution and 2) global coverage. On the Moon, only one sensor has both characteristics, the Narrow Angle Camera onboard NASA's Lunar Reconnaissance Orbiter (LRO). Since LRO's launch in 2009, NAC covered the entire surface of the Moon multiple times, returning more than 1.6 million optical images with a spatial resolution ranging from ~0.5 to ~1.5 m/pixel. These NAC images can be retrieved from the Planetary Data System (PDS) in a large variety of formats and processing levels. While there are many lunar global datasets available, the NAC image stack is the only data with spatial resolutions sufficient for the detection of small geomorphological and human-made objects. For this study, we used the uppermost layer of the pre-calibrated pyramid tif files (ptifs) that can be retrieved from the LROC image archive. These images provide the full spatial resolution (in the top layer), but are reduced in size, as they have been reduced to 8bit. The reduced size allows for optimized file download, handling, and processing times.

In a first step we collected all available NAC imagery over the Apollo 15 and 17 landing sites with spatial resolutions higher than 0.8 m/pixel. We then tiled these images in patches of 64x64 pixels, resulting in 255,000 NAC tiles. This dataset had a stride of thirty-two pixels, so that each pixel was included, on average, in four images. All used patches cover an area of approximately sixty-four million square meters and include the Apollo 15 and 17 landing sites.

\subsection{Model}

For our unsupervised data distribution learner, we used a variational autoencoder (VAE). [13, 14]
This autoencoder has an encode (decode) module layers eight layers deep that alternated
between convolution (transposed convolution) and batch norm layers.
By the end of the encoding layer, the data undergoes a learned transformation from
its native 3 channels x 64 pixels x 64 pixels image space size to 128 filters each of size 16 pixels x 16 pixels. 
We used a default informational bottleneck latent code dimension size of $2^6$, but varied this down to $2^3$ and up to $2^{11}$ to control model capacity, as well as training time, inference time and memory size, between experiment runs.
We used the reparametrization trick to allow differentiability through the stochastic sampling process of the VAE.
The loss bound that we sought to minimize included the calculation of the KL divergence between the distribution of latent codes and a multivariate Gaussian prior with dimension that is the size of the informational bottleneck latent code dimension.
We also used the approach of the beta-VAE literature [12]
to upweight the divergence term of the loss to promote disentangled learned representations. 
The anomaly score of an image relative to our trained model is the $L_2$ reconstruction loss between the image and the VAE's reconstruction of the image. 
A promising anomaly score, not yet fully evaluated, is the norm of the $\mu$ vector in the VAE's code of that image, which can be seen as a distance of the image's latent code from the mode of the latent code distribution, and hence a quantitative degree of anomaly-ness.

\subsection{Metrics, Evaluating Lunar Surface Anomalies}
For evaluating anomaly detection on the lunar surface, we took NAC image patches with human-made artifacts as positive anomalous samples for our experiments, and all image samples with no known human-made artifacts negative non-anomalous samples. 
Examples of these human-made artifacts in our study were lunar lander modules left from the American Apollo 15 and 17 missions. 
For model evaluation metrics, we used 
the average precision (AP) of our model's precision recall curves,
precision at total recall (PaTR), defined as
$$\textup{PaTR} = \textup{precision at the model's threshold where the model returns all positive samples},$$ 
and efficiency over manual vetting (EoMV), defined as  
$$\textup{EoMV} = \frac{\textup{PaTR}}{\textup{precision of the naive random anomaly scorer}}.$$ 
The motivation for EoMV is from considering it as the fraction of images needed to be vetted by a human using a random search over the ratio of images needed to be vetted by a human using our model to present images by descending anomaly score order.

\subsection{Software, Hardware}

We used PyTorch for training, JupyterLab with Python to coordinate the experiments, and the seaborn statistical visualization python package to view and plot results. 
We used an NVIDIA GeForce GTX 1070 and Intel Core i7 system with 512 GB SSD for training and validation. 

\section{Experiments}

For our first set of experiments, we sought to verify our proposed method using a toy dataset containing 200 image patches around the Apollo 17 landing site, containing four positive lunar technosignature patches. 
For our second set of experiments, we tuned our model architecture and hyper-parameters on a validation set containing the Apollo 17 landing site. We used a dataset of 2000 image patches around the Apollo 17 landing site with six positive patches. More positive patches were obtained over our first experiment through a finer vertical striding of the dataset creation. 
For our third set of experiments, we validated the performance of our method on a test set containing a new, unseen landing site, Apollo 15, with no tuning of model architecture or hyper-parameters on this dataset. We used a dataset of 8000 image patches around Apollo 15 landing site, which contained ten positive technosignature patches. The model architecture and all other model and optimizer hyper-parameters were learned one the train set above, and evaluated on our train set in a one-shot fashion, with results reported below. Extra care was taken to avoid hyperparameter leakages between optimizing our approach on the train set and evaluating performance on our train set, due to the acutely limited nature of human lunar technosignatures.

\section{Results}

\subsection{Lunar VAE Landing Site Validation}
\begin{figure}[h]  
    \centering
    \begin{minipage}[]{4cm}
        \centering
        \includegraphics[scale=0.33]{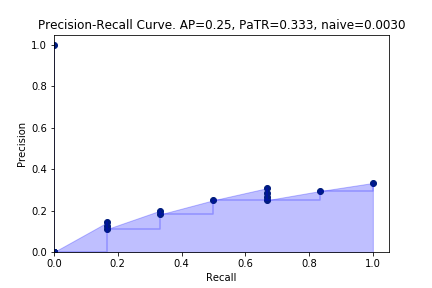}
    \end{minipage}
    \hspace{1cm}
    \begin{minipage}[]{4cm}
        \centering
        \includegraphics[scale=0.33]{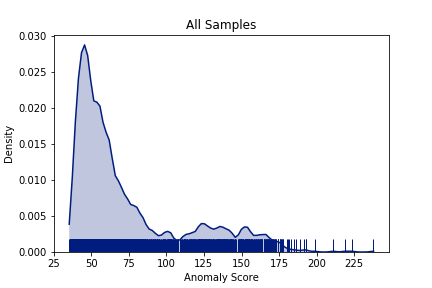}
    \end{minipage}
    \begin{minipage}[]{4cm}
        \centering
        \includegraphics[scale=0.33]{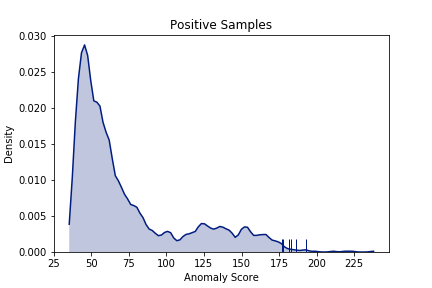}
    \end{minipage}
    \caption{Validation set precision-recall curve with metrics (left) and the distribution of anomaly scores of all samples on the horizontal axis (middle) versus only positive samples on the horizontal axis (right).}
    \label{fig:validation_figures}    
\end{figure}

For our smallest dataset containing 200 patches, our method achieved an AP and PaTR of 1.00 with minimal tuning by returning all positive samples before all negative samples.
Hence we moved to a more challenging and diverse validation dataset. 
For our larger validation dataset of 2000 patches,
in Figure \ref{fig:validation_figures} we plot the precision-recall curve with metrics for validation on our Apollo 17 train set, and the distribution of anomaly scores of all samples on the horizontal axis versus only positive samples on the horizontal axis. 
We achieve an AP of 0.25, PaTR of 0.333, resulting in an EoMV of $\frac{0.333} {0.003}$ = 111.
Note the unusual shape of the PR curve, which was due to a clustering of the positive sample anomaly scores into one mode. 

\subsection{Lunar VAE Landing Site Test}

\begin{figure}[h]  
    \centering
    \begin{minipage}[]{4cm}
        \centering
        \includegraphics[scale=0.33]{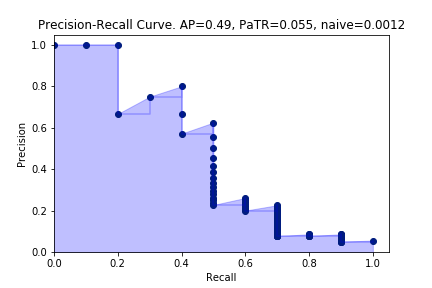}
    \end{minipage}
    \hspace{1cm}
    \begin{minipage}[]{4cm}
        \centering
        \includegraphics[scale=0.33]{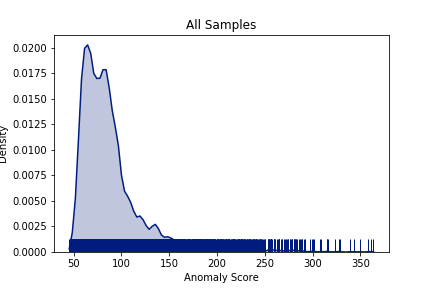}
    \end{minipage}
    \begin{minipage}[]{4cm}
        \centering
        \includegraphics[scale=0.33]{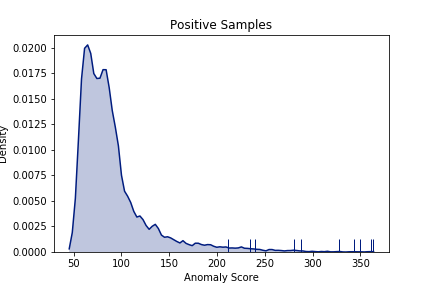}
    \end{minipage}
    \caption{Test set precision-recall curve with metrics (left) and the distribution of anomaly scores of all samples on the horizontal axis (middle) versus only positive samples on the horizontal axis (right).}
    \label{fig:test_figures}    
\end{figure}

In Figure \ref{fig:test_figures} we plot the precision-recall curve with metrics for validation on our Apollo 17 train set, and the distribution of anomaly scores of all samples on the horizontal axis versus only positive samples on the horizontal axis. 
We achieve an AP of 0.49, PaTR of 0.055, and an EoMV of $\frac{0.055} {0.0012}$ = 45.8. We note the high left hand portion of the PR curve, which was due to the top three returned samples being positive samples. 

\subsection{Evaluation of Learned Features} 
\begin{figure}[h]  
    \centering
    \begin{minipage}[]{4cm}
        \centering
        \includegraphics[scale=0.25]{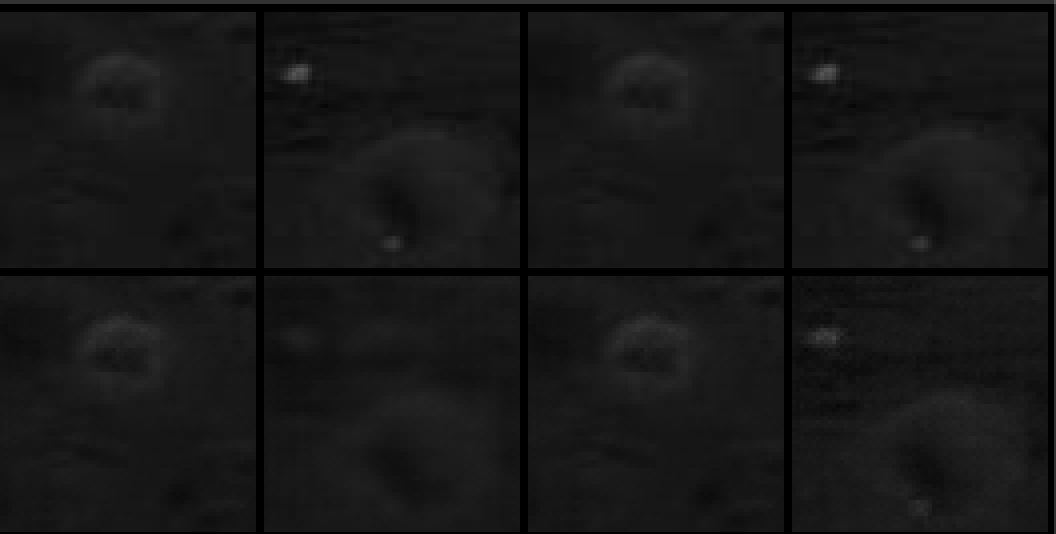}
    \end{minipage}
    \hspace{2cm}
    \begin{minipage}[]{4cm}
        \centering
        \includegraphics[scale=0.175]{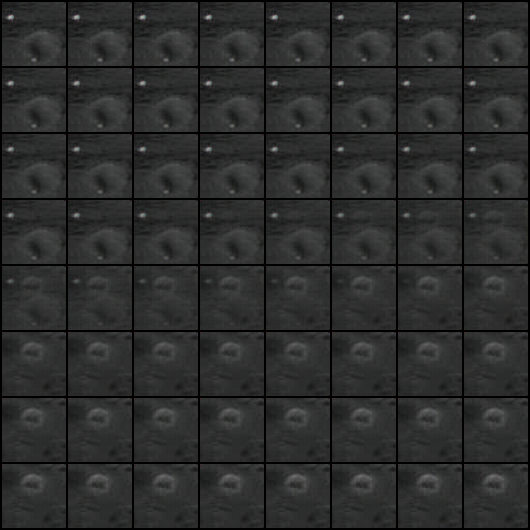}    
    \end{minipage}
    \caption{Sample reconstructions of two images (left) and a sample walk through the learned latent variable space (right).}
    \label{fig:evaluation_learned_features}    
\end{figure}

In Figure \ref{fig:evaluation_learned_features}, we plot two different input images each presented twice (top row) with their corresponding reconstructed images through the VAE's forward pass (bottom row). 
We note the stochastic nature of VAE generation. 
In the same figure we plot a sample walk through the learned latent variable space of our trained model on the Apollo 17 landing site. 
We consider the relative quality of reconstruction and the smoothness of interpolation between the two data samples at upper-left and lower-right hand corners as evidence that the unsupervised generative model was able to learn a good representation of the underlying lunar dataset, in addition to the performance of the learned distribution for technosignature detection. 


\section{Conclusion}


In this work we show that modern data-driven machine learning techniques can be successfully applied on lunar surface data to learn, in an unsupervised way, sufficiently good representations of the distribution of lunar surface data to enable lunar technosignature detection.
In particular we have trained an unsupervised distribution learning model to find the landing module of the Apollo 15 landing site in a testing dataset, with no specific model or hyperparamter tuning . Good data density estimation has myriad applications in lunar and space sciences, including finding known missions with unknown landing sites, discovering non-publicly disclosed landing sites, either by governmental or non-governmental organizations, technosignatures in other signal domains, locating lunar resources for future space flight and colonization, locating new impact craters or lunar surface reshaping, e.g. when applying this to temporal stacks of images, and deciding the importance of unlabeled samples to send back from power- and bandwidth-constrained missions. We hope this current work stimulates and enables future work towards these goals.


\newpage  

\section*{References}
\small
[1] N. Technosignatures Workshop Participants, NASA and the Search for Technosignatures: A Report from the NASA Technosignatures Workshop,
arXiv e-prints (2018) arXiv:1812.0868

[2] J. C. Tarter, The evolution of life in the universe: are we alone?, Proceedings of the International Astronomical Union 2 (14) (2006) 

[3] R. N. Bracewell, Communications from Superior Galactic Communities,
186 (1960) 670-671. doi:10.1038/186670a0.

[4] R. A. Freitas, F. Valdes, The search for extraterrestrial artifacts (SETA), Acta Astronautica 12 (12) (1985) 1027 

[5] P. C. W. Davies, R. V. Wagner, Searching for alien artifacts on the moon, Acta Astronautica 89 (2013) 261

[6] J. Haqq-Misra, R. K. Kopparapu, On the likelihood of non-terrestrial artifacts in the Solar System, Acta Astronautica 72 (2012) 15

[7] He, K., Zhang, X., Ren, S., \& Sun, J. (2016). Deep residual learning for image recognition. In Proceedings of the IEEE conference on computer vision and pattern recognition (pp. 770-778).

[8] Kingma, D. P., \& Welling, M. (2013). Auto-encoding variational bayes. arXiv preprint arXiv:1312.6114.

[9] Kingma, D. P., \& Welling, M. (2019). An Introduction to Variational Autoencoders. arXiv preprint arXiv:1906.02691.

[10] Kingma, D. P., \& Dhariwal, P. (2018). Glow: Generative flow with invertible 1x1 convolutions. In Advances in Neural Information Processing Systems (pp. 10215-10224).

[11] Reed, S., van den Oord, A., Kalchbrenner, N., Colmenarejo, S. G., Wang, Z., Chen, Y., ... \& de Freitas, N. (2017, August). Parallel multiscale autoregressive density estimation. In Proceedings of the 34th International Conference on Machine Learning-Volume 70 (pp. 2912-2921). JMLR. org.

[12] Higgins, I., Matthey, L., Pal, A., Burgess, C., Glorot, X., Botvinick, M., ... \& Lerchner, A. (2017). beta-VAE: Learning Basic Visual Concepts with a Constrained Variational Framework. ICLR, 2(5), 6.



\end{document}